# Quality Classification of Defective Parts from Injection Moulding

Adithya Venkatadri Hulagadri
School of Computer Science and Engineering
& Nanyang Business School,
Nanyang Technological University

Prof Mahardhika Pratama
School of Computer Science and Engineering,
Nanyang Technological University

Dr Edward Yapp Kien Yee
Singapore Institute of Manufacturing Technology

Andri Ashfahani
School of Computer Science and Engineering,
Nanyang Technological University

Lee Wen Siong,
School of Computer Science and Engineering,
Nanyang Technological University

*Abstract -* This report examines machine learning algorithms for detecting short forming and weaving in plastic parts produced by injection moulding. Transfer learning was implemented by using pretrained models and finetuning them on our dataset of 494 samples of 150 by 150 pixels images. The models tested were Xception, InceptionV3 and Resnet-50. Xception showed the highest overall accuracy (86.66%), followed by InceptionV3 (82.47%) and Resnet-50 (80.41%). Short forming was the easiest fault to identify, with the highest F1 score for each model.

## 1 INTRODUCTION

Injection Moulding is a popular method of manufacturing plastic and metal products by injecting molten materials into moulds at high pressure. This process is subjected to several random and systematic variations that sometimes lead to faults in the products that may be difficult to identify rapidly at large scales. This report focuses on data collected from the SIMTech Model Factory's Injection Moulding Machine that produces plastic parts for perfume cartridges. Two types of faults were examined – short-forming and weaving – and classified automatically by several machine learning algorithms.

Short-forming leads to smaller-than-required parts and occurs when the molten material does not cover the entire mould, typically occurring when either the holding pressure or injection speed of the molten material is too low. Weaving generally occurs due to inconsistent cooling of the molten material and may lead to lower durability. The models were therefore used for classification of parts into 3 categories: "Good", "Short-forming" and "Weaving".

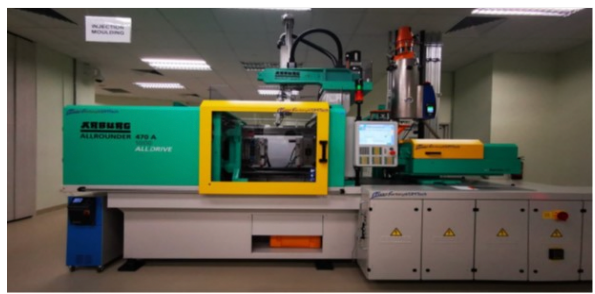

**Figure 1: Injection Moulding Machine**

This report examines several baseline Machine Learning algorithms for image classification and compares their effectiveness in classifying parts from injection moulding. Models examined include Xception, Resnet-50 and InceptionV3.

## 2 DATA COLLECTION

The data consisted of 25 experiments based on various settings of the injection moulding machine. The labels were then assigned by human supervisors at the SIMTech Model Factory based on 3 classes: Good, Weaving and Short forming.

Each image in the dataset was precisely 150 by 150 pixels and was registered with 3 channels –

red, blue and green. The number of images from each class are summarised below:

| Class | Number of Images |
|---|---|
| Good | 148 |
| Weaving | 147 |
| Short forming | 199 |
| Total Dataset | 494 |

**Figure 2: Dataset Summary**

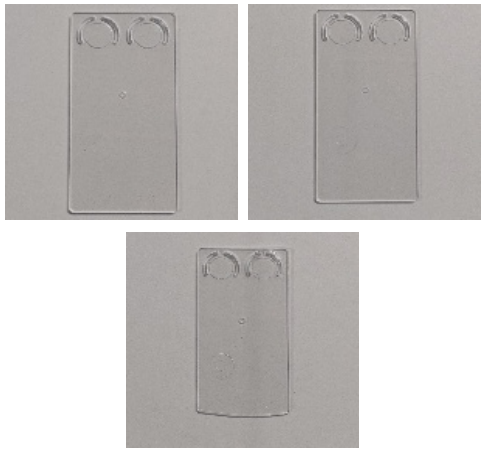

**Figure 3: Good parts, Weaving and Short forming respectively**

## 3 MODELS USED FOR TRANSFER LEARNING

Several benchmark models were used for classification, using a test-train split with 80 percent training data and 20 percent testing data. These models were implemented in python using the TensorFlow library and the pretrained models made available by the authors of the original publications.

Transfer learning was implemented by customising the input layer of each pretrained model to accept the new input photo size, and a custom classifier head was added to predict 3 classes instead of 1000 in the Imagenet dataset [5]. The preloaded weights of the remaining layers were retained and finetuned.

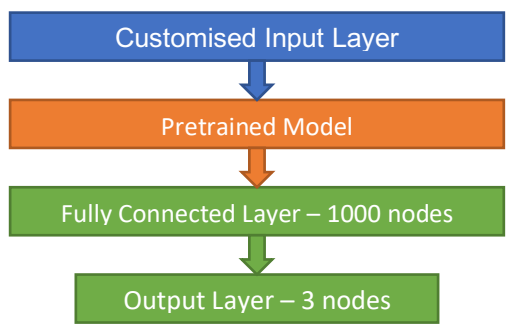

**Figure 4: Finetuning architecture**

The models used for finetuning were chosen for being sufficiently complex to model the Imagenet dataset and are listed below:

### 1. Xception

Deep Convolutional Neural network inspired by Inception V3 and is known to slightly outperform it [1] on the Imagenet dataset.

### 2. Inception V3

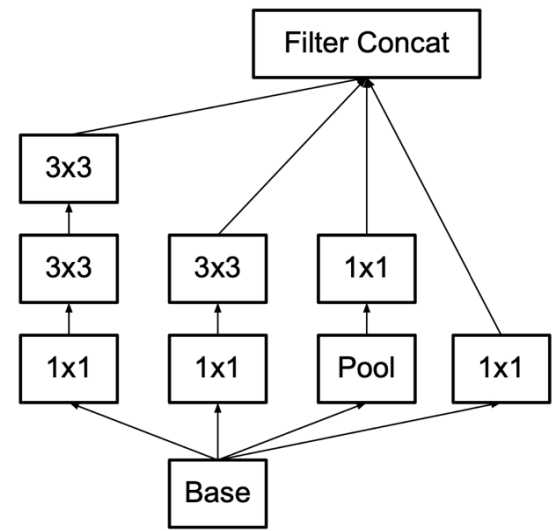

**Figure 5: Inception V3 filter architecture**

A deep convolutional neural network model using several units of "filters" and pooling layers [2].

### 3. Resnet-50

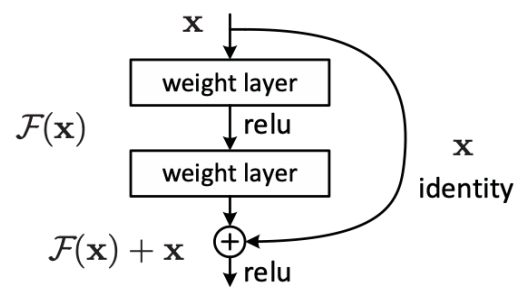

**Figure 6: Resnet Cell**

Solved the problem of vanishing gradients in very deep networks by using skip connections [3].

## 4 RESULTS USING TEST-TRAIN SPLIT

| Model | Best Validation Accuracy | Pretrained Model Size [4] |
|---|---|---|
| Xception | 0.86598 | 88 MB |
| Inception V3 | 0.82474 | 92 MB |
| Resnet-50 | 0.80412 | 98 MB |

Figure 7: Accuracy Results

| Xception | | Predictions | | | |
|---|---|---|---|---|---|
| | | Good | Weaving | Short forming | Recall |
| Actual | Good | 20 | 6 | 3 | 0.6897 |
| | Weaving | 1 | 27 | 1 | 0.9310 |
| | Short forming | 2 | 0 | 37 | 0.9487 |
| | Precision | 0.8696 | 0.8182 | 0.9024 | |

Figure 8: Confusion Matrix for Xception

| Inception V3 | | Predictions | | | |
|---|---|---|---|---|---|
| | | Good | Weaving | Short forming | Recall |
| Actual | Good | 18 | 7 | 4 | 0.6207 |
| | Weaving | 3 | 24 | 2 | 0.8276 |
| | Short forming | 1 | 0 | 38 | 0.9744 |
| | Precision | 0.8182 | 0.7742 | 0.8636 | |

Figure 9: Confusion Matrix for Inception V3

| Resnet-50 | | Predictions | | | |
|---|---|---|---|---|---|
| | | Good | Weaving | Short forming | Recall |
| Actual | Good | 23 | 3 | 3 | 0.7931 |
| | Weaving | 5 | 21 | 3 | 0.7241 |
| | Short forming | 2 | 3 | 34 | 0.8718 |
| | Precision | 0.7667 | 0.7778 | 0.8500 | |

Figure 10: Confusion Matrix for Resnet-50

| | Good | Weaving | Short forming |
|---|---|---|---|
| Xception | 0.7693 | 0.8710 | 0.9250 |
| Inception V3 | 0.7059 | 0.8000 | 0.9157 |
| Resnet-50 | 0.7797 | 0.7500 | 0.8608 |

Figure 11: F1 Scores

## 5 CONCLUSION

The results reveal that image models can be useful to detect faults in injection moulding, including weaving, which is difficult to identify with the naked eye.

All the models tested performed the best at detecting short forming, with the highest precision, recall and F1 score. This is expected, as it is easy to manually distinguish such parts with the naked eye.

Xception and Inception V3 performed better at detecting Weaving than good parts (with a higher F1 score for both models) while the opposite was seen in Resnet-50. Weaving was detected with a lower F1 score than for short forming. This could be attributed to the fact that weaving is difficult to identify visually as the fault is mainly internal, and only limited signs are observed on closer examination.

Given the limited usefulness of feeding images to detect subliminal faults, it may be necessary to design a dedicated neural network architecture to incorporate both images and sensor data which could carry information on internal faults.

## ACKNOWLEDGMENT

I would like to convey my gratitude to Prof Mahardhika Pratama of NTU and Dr Edward Yapp of SIMTech for supervising this project and providing valuable guidance and industry exposure. I am also grateful for the efforts of my collaborators: Andri Ashfahani provided valuable technical assistance in processing the data and Lee Wen Siong was essential in collecting the image data from the injection moulding machine.

Special thanks to SIMTech, for providing access to the injection moulding machine at the Model Factory, as well as to the team there for their assistance and helpful suggestions.

I would like to acknowledge the funding support from Nanyang Technological University – URECA Undergraduate Research Programme for this research project.